\documentstyle[emulateapj,psfig,apjfonts]{article}
\submitted{Received 2003 -- --; Accepted: 2003 -- --}

\def\gs{\mathrel{\raise0.35ex\hbox{$\scriptstyle >$}\kern-0.6em
\lower0.40ex\hbox{{$\scriptstyle \sim$}}}}
\def\ls{\mathrel{\raise0.35ex\hbox{$\scriptstyle <$}\kern-0.6em
\lower0.40ex\hbox{{$\scriptstyle \sim$}}}}

\makeatother

\lefthead{Smail et al.}
\righthead{{\it Chandra}/SCUBA sources around HzRGs}

\begin{document}

\title{{\it Chandra} Detections of SCUBA Galaxies around High-$z$ Radio Sources}

\author{
Ian Smail,\altaffilmark{1} C.\,A.\ Scharf,\altaffilmark{2}
R.\,J.\ Ivison,\altaffilmark{3} J.\,A.\ Stevens,\altaffilmark{3}
R.\,G.\ Bower\altaffilmark{1} \& J.\,S.\ Dunlop\altaffilmark{4}}

\altaffiltext{1}{Institute for Computational Cosmology, University of Durham, South Road,
        Durham DH1 3LE UK}
\altaffiltext{2}{Columbia Astrophysics Laboratory, Columbia University, MC5247, 550 
West 120th St., New York, NY 10027, USA}
\altaffiltext{3}{Astronomy Technology Centre, Royal Observatory, 
        Blackford Hill, Edinburgh EH9 3HJ UK}
\altaffiltext{4}{Institute for Astronomy, University of Edinburgh, 
        Blackford Hill, Edinburgh EH9 3HJ UK}

\setcounter{footnote}{4}

\begin{abstract}
The most massive galaxies in the present day universe are the giant
ellipticals found in the centers of rich clusters. These have old,
coeval stellar populations, suggesting they formed at high redshift,
and are expected to host supermassive black holes (SMBHs).  The recent
detection of several high-redshift radio galaxies (HzRGs) at
submillimeter (submm) wavelengths confirms that indeed some massive
galaxies may have formed the bulk of their stellar populations in
spectacular dust-enshrouded starbursts at high redshift. In this letter
we compare sensitive {\it Chandra} X-ray images -- which identify
actively-fueled SMBHs -- and submm observations -- capable of detecting
obscured activity in luminous galaxies at high redshift -- of the
environments of three HzRGs. These observations exhibit overdensities of
X-ray sources in all three fields and a close correspondence between
the {\it Chandra} and SCUBA populations.  This suggests that both
substantial star formation and nuclear activity may be occuring in
these regions.  We identify possible pairs of {\it Chandra} sources
with each of two SCUBA sources, suggesting that their ultraluminous
activity may be triggered by the interaction of two {\it massive}
galaxies, each of which hosts an accreting SMBH.  The presence of two
SMBHs in the resulting remanent is predicted to produce a flattened
stellar core in the galaxy, a morphological signature frequently seen
in luminous cluster ellipticals.  Hence the confirmation of pairs of
{\it Chandra} sources within individual, luminous SCUBA galaxies would
provide additional evidence that these galaxies at $z\sim 2$--4 are the
progenitors of the giant elliptical galaxies found in clusters at the
present-day.
\end{abstract}


\section{Introduction}

%
%
\begin{figure*}[tbh]
\centerline{\psfig{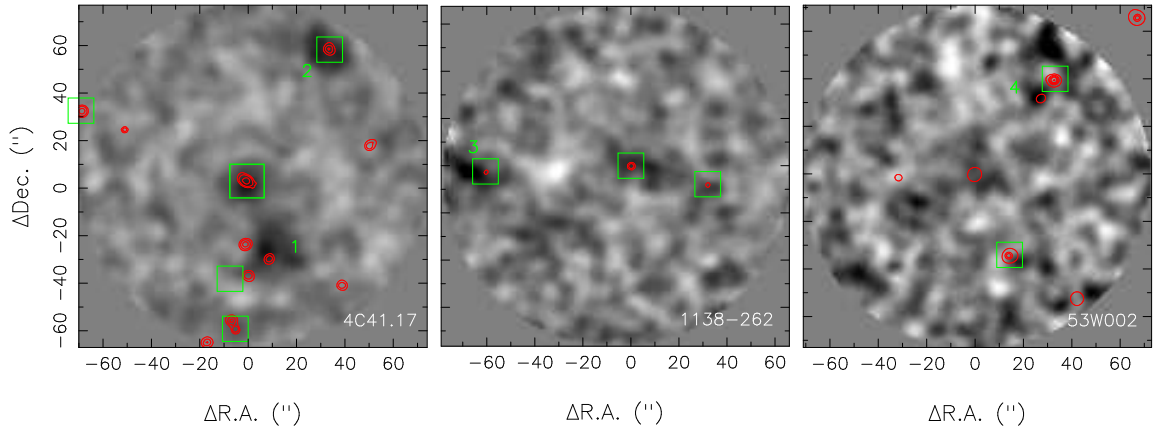}}
\caption{\small The 0.5--2\,keV {\it Chandra} X-ray images of the
regions around 4C\,41.17 (Scharf et al.\ 2003), MRC1138$-$262 (P02)
and 53W002 (White et al.\ 2003), each overlayed as a contour map on a
grayscale representation of the relevant 850$\mu$m SCUBA map from
Stevens et al.\ (2003).  We mark with \fbox{} the positions of the
hard-band sources detected in these fields.  Notice the X-ray
counterparts to four of the SCUBA sources in these field (numbered as in
Table~1), as well as the X-ray emission from the radio galaxies at the
center of each field.}
\end{figure*}

Studies of luminous ellipticals in high-density regions out to $z\sim
1$ suggest that their stellar populations are formed via a short burst
of star formation at high-$z$ (probably $z>2$), with little additional
star formation at $z<1$ (Bower, Lucey \& Ellis 1992; Ellis et al.\
1997; Stanford et al.\ 1998; Poggianti et al.\ 2001; Kelson et al.\
2001).  Further support for this conclusion comes from the apparent
lack of evolution in the metallicity of the intracluster medium out to
$z\sim 1$ (Tozzi et al.\ 2003).  The short dynamical times expected in
the densest regions at high-$z$ coupled with the intense star
formation necessary to form the stellar population of a luminous
elliptical suggest that this formation phase may resemble a classical
``monolithic'' collapse -- even within a hierarchical cosmogony
(Kauffmann 1996; Baugh et al.\ 1998).  A clear test of this paradigm
is to {\it directly} observe the epoch of elliptical formation in
high-$z$ protoclusters.  There are expected to be two distinct
observational signatures of this formation phase.

Firstly, the intensity of the star formation event and the high
metallicity of the stars in ellipticals today (as well as the
intracluster medium) all argue that this activity will
produce a large amount of dust. This will obscure the later
phases of the starburst and the most active systems, those with
star formation rates of $>10^2$--$10^3M_\odot$\,yr$^{-1}$, would
be classed as ultraluminous infrared galaxies (ULIRGs) with
far-infrared luminosities of $>10^{12}L_\odot$.  The sensitivity of 
submillimeter (submm) cameras such as SCUBA on the JCMT is sufficient
to detect these systems out to $z\gg 1$ (Chapman et al.\ 2003).

The second observational signature derives from the apparently close
correlation seen locally between the mass of central black holes and
the mass of the spheroid which hosts them (Kormendy \& Richstone 1995;
Magorrian et al.\ 1998; Gebhardt et al.\ 2000; Ferrarese \& Merritt
2000), which in turn suggests a tight linkage between the
formation of these components.  As the most massive spheroids, luminous
cluster ellipticals are expected to host supermassive black holes
(SMBH), $>10^9$M$_\odot$.  Thus many young ellipticals may also exhibit
signatures of nuclear activity as their SMBH will be actively fueled
and growing during the dusty starburst phase, and should be detectable in
deep X-ray imaging.

To investigate the formation of the elliptical galaxy population of
rich clusters using these observational tools we need to
identify the likely site of their formation: proto-clusters at $z>2$.
In standard cosmological models, massive galaxies at $z>2$ are
expected to strongly cluster in the deepest potential wells, onto which other
galaxies later accrete, assembling the rich clusters we see at $z\sim
0$. The strongly clustered population of luminous high-$z$ radio
galaxies (HzRG) are believed to host massive black holes and thus may
represent some of the most massive galaxies at these epochs (e.g.\ de
Breuck et al.\ 2000) which in turn could trace some of the highest
density environments. The association of luminous radio galaxies with
overdense structures at high-$z$ is supported by the identification of
modest excesses of companion galaxies around HzRGs in the
optical/near-infrared (e.g.\ R\"ottgering et al.\ 1996), X-ray
(Pentericci et al.\ 2002, P02; Fig.~1) and submm wavebands (Ivison et
al.\ 2000, I00; Stevens et al.\ 2003; Fig.~1).

We have brought together these two observational tools: submm and X-ray
imaging, to search for the formation phase of luminous elliptical
galaxies in putative high density environments around HzRGs.  We have
identified X-ray counterparts to the SCUBA sources in the fields of
three HzRGs.  We discuss the properties of these sources and the
information they provide on the formation of SMBHs and the early stages
of the growth of massive cluster elliptical galaxies.  We assume a
cosmology with $\Omega_m=0.27$, $\Omega_\Lambda=0.73$ and
$H_o=71$\,km\,s$^{-1}$\,Mpc$^{-1}$.

\section{Analysis and Results}

A search of the {\it Chandra} archive shows that three fields from the
SCUBA survey of HzRG by Stevens et al.\ (2003) have moderate or deep
X-ray imaging observations.  These are 4C\,41.17 ($z=3.79$),
MRC\,1138$-$262 ($z=2.16$) and 53W002 ($z=2.39$), with useable
integration times of 135.1\,ks, 32.5\,ks and 32.8\,ks respectively.
The {\it Chandra} observations of 4C\,41.17 are discussed in Scharf et
al.\ (2003), while those for MRC\,1138$-$262 are detailed in P02 and
Carilli et al.\ (2002), and 53W002 is presented in White et al.\
(2003). We refer the reader to these papers for more details of the
observations and their reduction.  Sources have been detected and
measured in these images using the {\sc ciao} software {\sc wavdetect}
algorithm with default settings. Hardness ratios (HR) are determined
as $(H-S)/(H+S)$ for a 0.5--2 keV soft band $(S)$, and 2--10 keV hard
band $(H)$.  To convert count rates to fluxes we assume a Galactic
$\log N(\rm H{\sc i})=21.0$ for all sources and a power law spectrum
with photon index $\Gamma=1.5$ and intrinsic $\log N(\rm H{\sc
i})=23.0$ for HR$>-0.5$ and $\Gamma=2.2$ with no intrinsic absorption
for HR$<-0.5$ (c.f.\ Tozzi et al.\ 2001). Luminosities are calculated
in the 0.3--10 keV rest frame using {\sc xspec}, errors assume
Poissonian count rate uncertainties.

In all three fields excesses of soft-band sources are seen: P02 show
that the MRC\,1138$-$262 field has a modest excess of soft-band
sources amounting to a factor of 50\% (in an $8'\times 8'$ field),
while White et al.\ (2003) estimate an overdensity of 2--3$\times$ in
soft-band sources around 53W002.  In the 4C\,41.17 field we find 11
soft-band sources brighter than $S_{0.5-2}=0.2\times
10^{-15}$\,erg\,s$^{-1}$\,cm$^{-2}$ within the SCUBA field (Fig.~1,
2.1$'$, 1\,Mpc diameter), compared to a blank field expectation of
$\sim 2$ (Rosati et al.\ 2002).  Thus we have a factor of 5--6$\times$
overdensity of soft-band sources in the vicinity of 4C\,41.17 (there
is also a $2\times$ overdensity of hard-band sources in this field).

As described by P02, the X-ray sources in MRC\,1138$-$262 also display
a strongly filamentary pattern around the radio galaxy suggestive of
associated large scale structure.  A similar, but weaker filamentary
structure may extend to the south of 4C\,41.17 (Fig.~1).  The
existance of overdense structures around the HzRGs is supported by
follow-up of the X-ray sources in the MRC\,1138$-$262 field where at
least 2, and possibly up to 5, sources lie at the same redshift as the
HzRG (P02).  Similarly, for 53W002, White et al.\ (2003) identify at
least 3 X-ray bright AGN at the same redshift as the HzRG. On-going
spectroscopy of the 4C\,41.17 field has so far confirmed one {\it
Chandra} companion at $z=3.8$.

%
%
\begin{table*}
{\scriptsize
\centerline{\sc Catalog of SCUBA/{\it Chandra} Sources}
\begin{center}
\begin{tabular}{lcccccccccl}
\hline\hline
\noalign{\smallskip}
Source & R.A.\ & Dec.\ & S$_{850}$ & S$_{0.5-2}$ & S$_{2-10}$ & HR & $L_X(0.3-10)$ & $\alpha_{SX}^a$ & P$^b$ & Comment \cr
   &  (J2000) & (J2000) & (mJy) & \multispan2{ ~~(10$^{-15}$\,erg\,s$^{-1}$\,cm$^{-2}$) } & & (10$^{43}$\,erg\,s$^{-1}$) & &  & \cr
\hline
\noalign{\smallskip}
1)~ {\bf SMM\,J06509+4130} & 06 50 51.2 & +41 30 05 & 15.6$\pm$1.8 & ... & ... & ... & ... & 1.18 & ... & 4C\,41.17 HzRG850.1, I00 \cr
~~CXJ\,065051.3+412958  & 06 50 51.30  & +41 29 58.1   &    ...       & 0.26$\pm$0.09 & 0.93$\pm$0.40    & $-0.3$ & 3.2$\pm$1.0   & 1.28 &  0.004 & HzRG850.1/K3, I00 \cr 
~~CXJ\,065052.1+412951  & 06 50 52.09  & +41 29 51.2   &    ...       &  0.38$\pm$1.0 & 1.9$\pm$0.6   & $-0.1$  & 5.6$\pm$1.2  & 1.24 & 0.047 &  \cr 
~~CXJ\,065052.2+413004  & 06 50 52.19   & +41 30 04.5   &    ...       &  0.85$\pm$1.6 & 0.44$\pm$0.25   & $-0.8$  &13.0$\pm$2.0  & 1.25 & 0.221 & LR1, Lacy \& Rawlings (1996) \cr 
\noalign{\smallskip}
2)~ {\bf SMM\,J06509+4131} & 06 50 49.3 & +41 31 27 & 8.7$\pm$1.2 & ... & ... & ... & ... &1.18 &  ... & 4C\,41.17 HzRG850.2, I00 \cr
~~CXJ\,065049.1+413127  & 06 50 49.13  & +41 31 26.5 &    ...       &  0.43$\pm$0.11   & 3.3$\pm$0.8    & $+0.1$ & 8.0$\pm$1.2 & 1.18 & 0.016 & HzRG850.2/K1, I00 \cr  
\noalign{\smallskip}
3)~ {\bf SMM\,J11409$-$2629}       & 11 40 53.3  &  $-$26 29 11 & 7.4$\pm$1.5 & ... & ... & ... & ... & 1.07 & ... &  MRC\,1138$-$262, Stevens et al.\ (2003)\cr
~~CXJ\,1114052.8$-$262911  & 11 40 52.84 & $-$26 29 11.2 & ... & 5.1$\pm$0.7 & 5.2$\pm$1.7    & $-0.7$ & 4.2$\pm$0.4 & 1.07 & 0.009 & \#9 ($z\sim 2.16$), P02 \cr  
\noalign{\smallskip}
4)~ {\bf SMM\,J17142+5016}       & 17 14 12.1 & +50 16 02 & 5.6$\pm$0.9 & ... & ... & ... & ... & 1.14 & ... & 53W002 \#18 ($z=2.39$), S03a\cr
~~CXJ\,171412.0+501602  & 17 14 11.97  & +50 16 02.3 & ... & 0.48$\pm$0.24 & 3.1$\pm$1.6    & $~0.0$ & 4.6$\pm$1.5 &  1.14 & 0.002 & \cr 
\hline
\end{tabular}
\end{center}
\addtolength{\baselineskip}{-1.0mm}
$a$) The submm--X-ray spectral index (Fabian et al.\ 2000).  The values for the SCUBA galaxies assume all the X-ray sources contribute to the submm emission, while those quoted for
the individual {\it Chandra} sources assume all the submm emission
arises just from that source.\\
$b$) Likelihood of detecting the observed submm flux at the X-ray position by random chance.

}
\end{table*}

Submm maps with noise levels of $\sim 1.0$--1.5\,mJy are available of
these three fields from the 850$\mu$m SCUBA survey of Stevens et al.\
(2003), with the first results on 4C\,41.17 reported in I00.  These
cleaned maps also show modest overdensities of submm sources compared
to blank fields, with at least one example, SMM\,J17142+5016 in the
53W002 field, spectroscopically confirmed at the same redshift as the
radio galaxy (Smail et al.\ 2003a, S03a).  There are 3 $>4$-$\sigma$
submm sources around 4C\,41.17 (I00), 3 around MRC\,1137$-$262 (Stevens
et al.\ 2003) and 4 around 53W002 (S03a).  Using the flux limits for
the three fields and the compilation of SCUBA counts in Smail et al.\
(2002), we estimate that the HzRG fields show $2\times$
overdensities of SCUBA galaxies compared to the field, with
roughly 4--5 of the sources  likely to be unrelated field galaxies.

To compare the X-ray and submm emission in these fields we align the
two maps.  The astrometric precision of the 850$\mu$m maps is $\pm
3''$ rms and small shifts of less than this magnitude have been
applied to align the submm emission with the radio centroids for the
luminous HzRGs in these fields.  The X-ray images are similarly tied
to the radio frame with an absolute precision of 0.5$''$ rms using the
positions of several X-ray and radio bright AGN in each field.

Due to the extended nature of the submm emission in many of the SCUBA
sources in our fields (I00; Stevens et al.\ 2003) we choose to test
the association of {\it Chandra} sources with the SCUBA galaxies by
searching for statistical excesses of submm emission around the
precisely-located X-ray source.  To calculate these fluxes we use a
3$''$-radius aperture which represents the relative uncertainty of the
astrometry between the {\it Chandra} and SCUBA maps. We measure the
flux in $\sim 10^3$ randomly-placed apertures across the SCUBA map and
calculate the fraction of these which exceed that measured at the
positions of the {\it Chandra} sources.  This procedure provides a
likelihood for each association and allows us to identify
statistically significant ($>$98\% c.l.) associations between X-ray
sources and 4 out of the 10 SCUBA galaxies in these fields (we
ignore the central HzRG in each field each of which are detected in
both the X-ray and submm wavebands, Fig.~1).   It should be
noted that these associations may result either from a direct correspondence
between the sources in the two wavebands, or because both classes
populate the same structures around the HzRGs. We list in Table~1 the
details of these associations and briefly discuss the individual cases
below.

\noindent{\it 1) SMM\,J06509+4130} -- this is the very bright
($L_{bol}\sim 10^{13}L_\odot$), southern SCUBA source in Fig.~1a
(HzRG850.1 from I00).  This resolved submm source (70\,kpc diameter)
coincides with the soft-band {\it Chandra} source
CXJ\,065051.3+412958, which is identified as the $K=19.4$ very red
object HzRG850.1/K3 from I00.  A second soft-band {\it Chandra} source
to the South-East, CXJ\,065052.1+412951, is also associated with the
submm emission ($<$5\% chance of this being random coincidence). This
source has no obvious optical or near-IR counterparts in the imaging
discussed by I00 indicating $K>20.0$ and $R>26.0$.  A third, bright
soft-band {\it Chandra} source lies just outside the region of submm
emission to the East (Fig.~1a).  This source is coincident with a
red stellar object (LR1 from Lacy \& Rawlings 1996) with $UVR$ colors
consistent with either a foreground star or a $z\sim 3.8$ AGN (there
is a 6\% chance of an unrelated X-ray source falling this close to the
SCUBA position).  Finally, a hard-band source with a faint $K$-band
counterpart is detected 18$''$ to the South-East, although this is
unlikely to be directly related to the SCUBA galaxy.
 
\noindent{\it 2) SMM\,J06509+4131} -- this is the second, bright and
spatially-extended, SCUBA galaxy (HzRG850.2 in I00) in  
the 4C\,41.17 field (Fig.~1a) and is associated with the {\it Chandra} source
CXJ\,065049.1+413127.  The X-ray source is detected in both the hard
and soft-bands and is coincident with the K1 counterpart of
HzRG850.2 from I00.  K1 is one of a pair of faint, very red
galaxies separated by $\sim 2''$.

\noindent{\it 3) SMM\,J11409$-$2629} -- this is the brightest submm
source in the vicinity of the $z=2.16$ radio galaxy MRC1138$-$262 
mapped by Stevens et al.\ (2003; Fig.~1b).  The {\it Chandra} source
CXJ\,1114052.8$-$262911 (\#9 from P02) is statistically associated
with the SCUBA source (Table~1).  P02 report that \#9 exhibits an Ly$\alpha$
emission-line excess, indicating that it probably lies
at $z\sim 2.16$, which would place this SCUBA galaxy in the striking
filament seen across this field.

\noindent{\it 4) SMM\,J17142+5016} -- this relatively faint SCUBA
galaxy has been identified by S03a with a weak AGN, \#18 from the catalog of
Keel et al.\ (2002).  The {\it Chandra}
source coincides exactly with the optical/near-IR counterpart and
hence we can unequivocally associate both the X-ray and luminous submm
emission with this galaxy which is spectroscopically-confirmed as a
companion to the radio galaxy at $z=2.39$.  A much brighter X-ray
source is detected 9$''$ (75\,kpc) to the North-West, coincident with
the broad-line $z=2.39$ AGN \#19 (Keel et al.\ 2002), although this
source is undetected in the SCUBA map (Fig.~1c).  There is a 2\% chance
that this source would fall within 9$''$ of the SCUBA galaxy by
random chance.

In summary, we detect X-ray counterparts to four luminous submm sources
in the fields of three $z=2.2$--3.8 HzRGs.  Moreover, one of these
X-ray sources has the same redshift as the radio galaxy in its vicinity
and a second is also likely to be at the same redshift as its
neighboring HzRG. These identifications confirm the presence of
luminous dust-obscured galaxies containing actively-fueled SMBHs in the
environments of radio galaxies at $z>2$.  We give their hardness ratios
and submm/X-ray spectral indices ($\alpha_{SX}$) in Table~1 and also
estimate their X-ray luminosities assuming they lie at the same
redshifts as the relevant HzRG.  For all four sources these X-ray
properties are consistent with obscured, Seyfert-like systems, $L_X\sim
10^{43}$--$10^{44}$\,ergs\,s$^{-1}$ with absorbing columns of $N(\rm
H{\sc i})\gs 10^{23}$\,cm$^{-2}$ (Alexander et al.\ 2003a; Fabian et
al.\ 2000).  We obtain a lower limit on the black hole masses in these
systems of $\gs 10^{7}$M$_\odot$ by assuming that they are accreting at
their Eddington limit, or $\sim 10^{8}$M$_\odot$ if their SMBHs are
accreting at a similar rate to local Seyferts.

\section{Discussion and Conclusions}

Our identifications of {\it Chandra} counterparts to the SCUBA galaxies
in these fields are statistical in nature and hence the possibility
exists that the two populations are simply tracing the same large-scale
structures, rather than being {\it directly} related.  However, the
relatively small spatial scale of the associations,$\ls 100$\,kpc,
suggests that there is a direct relationship and so the following
discussion proceeds on that assumption.

The rate of detection of SCUBA sources in our {\it Chandra} maps is
$\sim 40$\%, this is somewhat higher than the $\sim 5$--10\% detection
rate reported from comparable-depth {\it Chandra} and {\it XMM-Newton}
observations (Barger et al.\ 2001; Ivison et al.\ 2002; Almaini et al.\
2003; Waskett et al.\ 2003) and is similar to that achieved in the
far-deeper, 2-Ms {\it Chandra} integration discussed by Alexander et
al.\ (2003), who detect four out of ten $\geq 5$\,mJy SCUBA sources
above a flux limit of $S_{0.5-8}\geq 1\times
10^{-15}$\,erg\,s$^{-1}$\,cm$^{-2}$.  Our high detection rate could
either reflect enhanced X-ray emission from SCUBA galaxies in these
environments, or may instead simply point to an increase in the
detection rate of $\gs 5$\,mJy SCUBA sources just below the flux limits
of the shallower surveys, e.g.\ $S_{0.5-8}\sim 3\times
10^{-15}$\,erg\,s$^{-1}$\,cm$^{-2}$ for the Almaini et al.\ (2003)
study (Manners et al.\ 2003).

We also identify a significant overdensity of {\it Chandra} sources
within a 1-Mpc region around the radio source in the 4C\,41.17 field,
in both the soft and hard-bands.  Similar overdensities have been
previously reported around MRC\,1138$-$262 (P02) and 53W002 (White et
al.\ 2003), suggesting that the environment of HzRGs at $z>2$ are also
fertile ground for triggering AGN activity.  This activity appears to
be linked to the excess of luminous submm sources found in these regions,
as can be seen if we measure the mean submm flux of {\it all} {\it
Chandra} sources within our SCUBA maps.  We estimate a mean flux of
$2.6\pm 0.7$\,mJy from the 15 soft-band sources (excluding the 3 radio
galaxies). This flux is roughly twice that of blank-field {\it Chandra}
sources at similar X-ray flux limits (Almaini et al.\ 2003; Barger et
al.\ 2001; Waskett et al.\ 2003), although the result is only
significant at the $\sim 2$-$\sigma$ level due to the small sample
size. 

Two of the SCUBA sources detected by {\it Chandra} are particularly
noteworthy.  The first is SMM\,J06509+4130 (HzRG850.1 from I00), a very
luminous and spatially-extended submm source which is apparently
associated with two X-ray sources (separated by 11$''$ or 80\,kpc if at
$z=3.8$), with two other X-ray sources in close proximity.  A second
X-ray detected SCUBA source is SMM\,J17142+5016, which also has an
X-ray bright companion, 53W002\#19, at a projected distance of 80\,kpc
and a velocity offset of only 400\,km\,s$^{-1}$ (Keel et al.\ 2002).
Two similar pairs of {\it Chandra} sources associated with SCUBA
galaxies, with separations of 2--3$''$, have been reported by Alexander
et al.\ (2003a) from a sample of 10 SCUBA galaxies in the CDF-N.  Thus
from a combined sample of 21 SCUBA sources we have 3 (14\%) with pairs
of {\it Chandra} counterparts brighter than $S_{0.5-8}\sim
10^{-15}$\,erg\,s$^{-1}$\,cm$^{-2}$, this is $7\times$ higher than the
occurence of similar separation pairs in the general {\it Chandra}
population at this flux limit (9/460 or 2\%, Alexander et al.\ 2003b).

These SCUBA galaxies with pairs of {\it Chandra} counterparts are
remarkable systems -- suggesting that the ultraluminous SCUBA events
are triggered by the interaction between multiple(?) massive galaxies.
The activity in these paired {\it Chandra} sources implies that they
are currently interacting and their separations then suggests that the
pericentric passage probably occured within the last $\sim 100$\,Myrs,
consistent with the expected life time of the starbursts in SCUBA
galaxies (Smail et al.\ 2003b).  Moreover, the presence of SMBHs in the
progenitors means they have either been formed on the timescale of the
current interaction, or these galaxies had sufficient prior activity to
build a SMBH (and by implication a massive spheroid), yet still retain
significant quantities of gas needed to power the current, obscured
starburst.

The final merger of these pairs of galaxies, probably within the next
100--200\,Myrs or by $z\sim 3.4$ for the 4C\,41.17 pair, will produce
a remanent which hosts two SMBHs.  The interaction of the two SMBHs is
predicted to produce a flattening of the stellar distribution in the
core of the remanent (Ravindranath, Ho \& Filippenko 2002).  This is a
structural feature common to bright elliptical galaxies, and also one
which is preferentially found in ellipticals in clusters (Laine et
al.\ 2003; Quillen, Bower \& Stritzinger 2000).  Hence the
identification of multiple {\it Chandra} sources within a modest
fraction of the SCUBA population is consistent with the expectation
that these galaxies will evolve into luminous ellipticals at the
present-day and that the environments we find them in at high-$z$ are
regions which will go on to become the cores of rich clusters.
However, the relatively modest mass estimates we derive for these
SMBH, $\gs 10^7$--$10^8$M$_\odot$, compared to the $\sim
10^9$M$_\odot$ SMBHs found in bright ellipticals today gives scope for
subsequent growth of the SMBHs in these systems during the final
stages of the merger.

The success rate for identifying {\it Chandra} counterparts to SCUBA galaxies  
in our X-ray images suggests that X-ray observations may
provide a useful route to localise the submm-emitting galaxies in
these regions.  This would normally be achieved through sensitive
radio observations (e.g.\ Ivison et al.\ 2002), but the presence of
the very bright radio source in these fields restricts the depth that
can be achieved due to dynamic-range limitations.

We summarise the main results of this work.  We study moderate depth
X-ray observations of luminous submm galaxies in the fields of three
luminous radio galaxies at $z=2.2$--3.8.  We identify a significant
overdensity of X-ray sources around the most distant target, 4C\,41.17
at $z=3.8$.  We also identify possible X-ray counterparts to 4 submm
galaxies in these fields (as well as the 3 submm-luminous central radio
galaxies).  These associations either indicate a direct relationship
between the {\it Chandra} and SCUBA sources, or simply that both
populations are tracing the same large-scale structures.  Spectroscopic
and narrow-band imaging suggests that two of these {\it Chandra}
sources lie at the same redshifts as their neighbouring HzRGs, hence
our identification of these objects as submm sources extends the number
of dusty, ultraluminous galaxies likely to reside in structures around
radio galaxies at $z>2$. We also highlight the properties of two
of these submm galaxies, one appears to be associated with at least
{\it two} {\it Chandra} sources, while the second has a {\it Chandra}
counterpart and also a bright X-ray companion in close proximity.
These results suggest that the ultraluminous activity in some fraction
of the SCUBA population may result from the interaction of pairs of
massive galaxies, each of which can be identified by the X-ray emission
from their central supermassive black holes.  The presence of two SMBHs
in the resulting merger remanent is predicted to produce a clear
structural signature, a flattened core, in the stellar profile.  Such
features are commonly observed in luminous elliptical galaxies in
clusters, suggesting a link between SCUBA galaxies at $z>2$--4 and the
population of luminous elliptical galaxies seen in clusters today.

\acknowledgments
We thank Dave Alexander, Omar Almaini, Wil van Breugel, Arjun Dey and
Bill Keel for help and useful conversations.  We also
thank an anonymous referee for suggestions which improved the
presentation of this work. IRS acknowledges support
from the Royal Society and the Leverhulme Trust, JAS, RGB and JSD
acknowledge support from PPARC.  This research was supported by
NASA/{\it Chandra} grant SAO G02-3267X.  This work was made possible
by the exceptional capabilities of the NASA/{\it Chandra} observatory,
and the dedicated support of the CXC.

\end{document}